\documentstyle[prl,floats,preprint,aps]{revtex}
\tighten

\begin{document}
\draft

\preprint{\vbox{\hbox{MIT-CTP-2499}
\hbox{hep-ph/9601296}}}

\title{Supernatural Inflation}

\author{Lisa Randall, Marin Solja\v{c}i\'{c}, and Alan H. Guth}

\address{Laboratory for Nuclear Science and Department of Physics \\
Massachusettts Institute of Technology \\ Cambridge, MA 02139}

\date{January 18, 1996}
\maketitle
\begin{abstract}
Most models of inflation have small parameters, either to
guarantee sufficient inflation or the correct magnitude of the
density perturbations. In this paper we show that, in
supersymmetric theories with weak scale supersymmetry breaking,
one can construct viable inflationary models in which the
requisite parameters appear naturally in the form of the ratio of
mass scales that are already present in the theory.  Successful
inflationary models can be constructed from the flat-direction
fields of a renormalizable supersymmetric potential, and such
models can be realized even in the context of a simple GUT
extension of the MSSM.  We evade naive ``naturalness" arguments
by allowing for more than one field to be relevant to inflation,
as in ``hybrid inflation" models, and we argue that this is the
most natural possibility if inflaton fields are to be associated
with flat direction fields of a supersymmetric theory. Such
models predict a very low Hubble constant during inflation, of
order $10^3$-$10^4$ GeV, a scalar density perturbation index $n$
which is very close to or greater than unity, and negligible
tensor perturbations.  In addition, these models lead to a large
spike in the density perturbation spectrum at short wavelengths.
\end{abstract}

\narrowtext 

\section{Introduction}

Inflationary models \cite{in} in general require small parameters
in the particle theory Lagrangian, to provide the flat potential
needed for sufficient inflation and for the correct magnitude of
density fluctuations.  The need for unmotivated small parameters
tends to weaken the credibility of a theory, so one hopes that
the origin of these parameters can be understood.  It is
conceivable, of course, that the explanation lies beyond our
present understanding, just as we presently have no accepted
explanation of why the Yukawa coupling of the electron is $2
\times 10^{-6}$, or why the weak scale lies 17 orders of
magnitude below the Planck scale.  Nonetheless, it would be
encouraging to find that the small parameters required by
inflation could be obtained from small parameters that are
already essential to the particle theory, so that no additional
small parameters are introduced.  Models in which the small
parameters arise as ratios of known particle physics mass scales
are particularly attractive
\cite{ADFKRS}.

{}From a field theoretical perspective, it is difficult to see
how flat direction fields can be present in a nonsupersymmetric
theory (with the exception of Goldstone bosons, considered in
Ref. \cite{natural}) given the effect of radiative corrections. 
We will assume therefore that the world is supersymmetric, and
ask whether an inflationary potential can arise naturally in the
context of the mass scales which we expect might be present. Some
natural candidates for these scales could be the Planck scale
$M_p \approx 10^{19}$ GeV, the GUT scale $M_{\rm G}
\approx 10^{16}$ GeV, the intermediate scale $M_I \approx
10^{11}$ GeV, and the supersymmetry breaking scale $m_{3/2}
\approx 1$ TeV.

In the context of supersymmetric models, an attractive scale for
the vacuum energy density during inflation would be set by the
intermediate scale, $M_I$.  This scale is very likely to be
present in a hidden sector model of supersymmetry breaking. It is
also the right energy scale for the potential associated with
moduli fields, which might be natural candidates for flat
directions.  The problem here, however, is that simple
dimensional analysis arguments (to be reviewed in Sec.~II) show
that density perturbations would generically be either far too
small \cite{rt,b1} or far too large, depending on assumptions. 
For the case of inflation driven by a single moduli field, the
dimensional arguments show that the requirements of sufficient
inflation and correct density perturbations imply that 1) the
variation of the inflaton (moduli) field during inflation is of
order $M_p$, and 2) the energy density during inflation is of
order $M^4$ where $M\approx 10^{16}$ \hbox{GeV}. An energy
density of order $M_I^4$ would produce density perturbations too
small by about ten orders of magnitude.  For the case of a
chaotic inflationary scenario, the variation of the inflaton is
again of order $M_p$, but in this case the density perturbations
are much too large unless the quartic coupling $\lambda$ is about
$10^{-12}$.

In this paper, we show that the argument that inflation at an
intermediate scale is untenable lacks sufficient generality, and
can evaporate if one drops the assumption that inflation is
driven by a single scalar field.  We describe a class of
two-field models, for which dimensional analysis estimates show
that 1) the variation of the inflaton is of order $M_I$ or less,
and 2) the energy density during inflation is of order $M_I^4$. 
We then go on to illustrate these ideas with models motivated by
supersymmetry with soft supersymmetry breaking. We will find that
these models not only solve the naturalness problem of obtaining
sufficiently many e-foldings of inflation, but also generate very
nearly the correct size of density perturbations based on the
parameters of supersymmetry breaking. We therefore refer to our
models as ``supernatural" inflation.

This model contains a similar structure to the ``hybrid"
inflation models, proposed by Linde and studied by Copeland,
Liddle, Lyth, Stewart, and Wands \cite{hybrid}.  The fact that
the standard dimensional naturalness arguments for the number of
$e$-foldings and for $\delta \rho/\rho$ do not apply, and that
the Hubble scale during inflation will be low was also clearly
recognized by these authors. Our point here is to emphasize that
the most natural scales for successful implementation of two
field inflation of the ``waterfall" type are the scales
associated with supersymmetry breaking and the Planck scale. 
Furthermore, our models more accurately reflect masses and
couplings associated with flat direction fields, and we will
motivate the parameters and potential we use by consideration of
flat directions in the \hbox{MSSM}.  Hybrid inflation in the
context of SUSY leads one to the interesting conclusion that the
Hubble scale during the inflation which established the density
perturbations might have been of order $10^3$--$10^4$ GeV, rather
than $10^{13}$ GeV.

In the following section, we present the general arguments for
why supersymmetry scales do not work in single field inflation
models. We then review the general idea of ``hybrid" or
``waterfall" \cite{hybrid} models, and show why the single-field
arguments do not apply to the two-field case.  In Sec.~III, we
present supernatural inflation models, in which we assume the
inflation sector consists of flat direction fields whose
potential is generated through supersymmetry breaking and
operators in the superpotential which do not permit the two
fields to be simultaneously flat.  We derive the constraints on
parameters consistent with the requisite number of $e$-foldings
and density perturbations.  We conclude in the final section.

\section{One vs. Two Field Inflation}

We begin this section by reviewing the ``standard" arguments for
why the inflaton in ``natural" inflationary models varies on the
scale $M_p$ and why the scale for the energy density should be
larger than the intermediate scale in inflationary models with a
single field.

For the purposes of these dimensional arguments, we first assume
the potential takes the form
\begin{equation}
{\cal V}=M^4 {\cal G}(\phi/f)\ ,
\end{equation}
where ${\cal G}$ is a bounded function of order unity.  Here we
have in mind for example a moduli {field,} with $M\approx M_I$. 
If we assume the slow roll equation of motion
$3H\dot{\phi}=-{\cal V'}$, where $H$ is the Hubble constant
during inflation, the number of $e$-foldings is
\begin{equation}\label{neqn}
N=\int H dt=\int d\phi {H \over \dot{\phi}}=- \int d\phi
{{\cal V} \over M_p^2 {\cal V'}}\approx-  {\Delta \phi \over
M_p} {f \over M_p} {{\cal G}\over {\cal G'}} \ .
\end{equation}
There are essentially two possibilities.  If ${\cal G}$ is a
bounded function, and ${\cal G'}$ is not very tuned to have very
flat sections, one is in the regime of what might be expected for
a moduli type field. In this case, the requirement of about 60
$e$-foldings of inflation favors $f$ of order $M_p$ and a change
in $\phi$ during inflation at least of order $M_p$. Even when
this is satisfied, some tuning of the potential is required.

The alternative possibility is that one is in a chaotic
\cite{chaotic} inflationary scenario, in which case ${\cal G}$
will be dominated by monomial behavior for sufficiently large
field, and ${\cal V'}\approx {\cal V} /\phi$.  In this case, $f$
is not defined, but one would still conclude $\Delta \phi
\approx M_p$.

Density fluctuations are also readily estimated under the assumed
form of the potential.  They are given by
\begin{equation}
{\delta \rho \over \rho}\approx {H^2 \over \dot{\phi}}\approx
{H^3 \over {\cal V'}}\approx \left({M \over M_p}\right)^2 {f
\over M_p} {{\cal G}^{3/2} \over {\cal G'}} \ .
\end{equation}
Assuming a potential of the moduli type, with ${\cal G}$ and
${\cal G'}$ of order unity and $f$ of order $M_p$, we find that
${\delta \rho / \rho} \approx \left({M / M_p}\right)^2$ favoring
$M \approx 10^{-3} M_p$. Detailed calculations might change $M$
by an order of magnitude or so, but it is clear that $M \approx
10^{11}{\rm \ GeV}\approx M_I$ is strongly disfavored.

In a chaotic scenario on the other hand, one would conclude that
the density fluctuations are too large unless there is a small
parameter. For example, a simple dimensional argument would lead
to the conclusion that for $V=\lambda \phi^4$, $\lambda \approx
10^{-12}$.  Without further motivation for these small numbers,
such a potential seems unlikely.

So one is led to the conclusion that it is difficult to naturally
obtain sufficiently many e-foldings and the correct magnitude of
density perturbations, without invoking either small numbers or a
new mass scale.

It is apparent, however, that there is a loophole in the above
argument.  From Eq.~(\ref{neqn}) it is clear that the constraints
on $f$ and $\Delta \phi$ during inflation arise because it is the
same potential ${\cal V}(\phi)$ that controls the inflation rate
$H$ and the speed of the inflaton field $\dot \phi$.  These
constraints can be avoided, therefore, if the energy density
during inflation is provided from some source other than the
scalar field which rolls and controls the ending of inflation.

The simplest way to implement this idea would be with two fields. 
This idea is essentially that first proposed by Linde
\cite{hybrid} as ``hybrid" inflation or ``waterfall" models. 
There are two fields $\psi$ and $\phi$. The first field, which we
call the inflaton, has a very flat potential. It starts at a
large field value, and slowly rolls (via its classical field
equations) to the origin. 

The second field, $\phi$, has a potential whose minimum is far
from the origin.  In most previous incarnations of hybrid
inflation, the scale of variation of this field is $M_I$, though
in our models the scale will be $M_p$.  When $\psi$ has large
field value, it gives a {\it positive} mass squared term in the
$\phi$ potential at the origin, so the classical field equations
{push} $\phi$ to the origin.  When $\psi$ gets sufficiently small
(of order $M_I$ or less in our models), the {mass squared} of
$\phi$ goes negative, and $\phi$ makes the transition from the
origin to $M_p$.

The key feature of this model is that the energy density during
inflation is dominated by the potential energy of the $\phi$
field at $\phi=0$.  There are no tunings in the $\psi$ potential
to get a small mass during inflation and a large mass afterwards,
since its mass is always small, as is its potential energy. 
Because $H$ depends on the value of $\phi$ and is not determined
by the field $\psi$, which acts as a switch to end inflation, the
naive estimates do not apply.

The second key feature of this model is that the ending of
inflation is controlled by when the $\phi$ mass squared at the
origin changes sign.  One can obtain a large number of
$e$-foldings with the variation of the inflaton field $\psi$ much
less than $M_p$.  Let us see this explicitly.

We assume a potential which takes the form
\begin{equation} \label{pot}
{\cal V}=M^4 {\cal G}(|\phi|/f)+g(|\phi|, |\psi|)+m^2 |\psi|^2 \ ,
\end{equation}
where $M$ and $f$ are to be determined, the function $g$ is the
term responsible for the $\psi$ dependence of the $\phi$ mass,
and $m$ is of order $m_{3/2}$.

We now have
\begin{equation}
N \approx \int d\psi {H \over \dot{\psi}}\approx - \int {H^2
d\psi \over m^2 \psi}\approx {M^4 \over m^2 M_p^2}
\ln\left({\psi_{init} \over \psi_{final}}\right)\  .
\end{equation}
Notice that the scale $M$ in the numerator is independent of the
mass and coupling of the $\psi$ field (in the limit that the
$\psi$ contribution to the energy density is small) so that the
previous arguments for one-field inflation no longer apply. 
Clearly for inflation to give several $e$-foldings requires only
that $\psi$ changes by an order of magnitude, and that $M^4
{~\lower0.6ex\hbox{\vbox{\offinterlineskip\hbox{$>$}\vskip1pt\hbox{$\sim$}}}~}
m^2 M_p^2$. No inflaton variation of order $M_p$ is required, and
so far, it seems $M\approx M_I$ could be a good choice.

Let us now consider density fluctuations under the same assumed
form for the potential. We find
\begin{equation}\label{6}
{\delta \rho \over \rho}\approx {H^2 \over \dot{\psi}}\approx
{H^3 \over m^2\psi}\approx{M^6 \over M_p^3 m^2 \psi} \ .
\end{equation}
The point is that the numerator $H^3$ has its scale set by the
$\phi$ potential energy while the denominator is determined by
the $\psi$ field. We construct a model so that $\psi$ at the end
of inflation is of the order $M_I$ or smaller.  If we also take
$M \approx M_I$, we find $\delta \rho /\rho \approx M_I/M_p$ or
bigger (rather than $(M_I/M_p)^2$ as was the case in single field
models).  Although the coupling between the fields can have a
coefficient which varies by many orders of magnitude, as does
$\psi$ in Eq.~(\ref{6}), the strong $M$ dependence of
Eq.~(\ref{6}) allows for agreement with the COBE constraint with
only a relatively small $M$ variation. This is very promising
from the perspective of relating inflation models to real scales
of particle physics. To answer the questions of how well these
ideas really work, and how constrained the parameters of the
models really are, requires a detailed investigation of
particular examples of these ideas. 

\section{ Supernatural Inflation }

We define Flat Direction Hybrid Inflation (FDHI) models as those
motivated by the properties of moduli fields or flat directions
of the standard model. 

In the first model, we assume the existence of a superpotential
which couples $\psi$ and $\phi$ but which is suppressed by a
large mass scale $M'$.  For standard model flat directions, such
higher dimension operators are to be expected, with $M'$ equal to
$M_p$, $M_G$, or some dynamical scale.  In the case of moduli
fields, it might be that this scale is of dynamical origin; one
can readily determine how the answer changes with the form of the
superpotential and the size of the mass scale. The example we
take is
\begin{equation}
W={\phi^2 \psi^2 \over 2 M'}\ .\label{supp} 
\end{equation}

We now need to specify the form of the supersymmetry breaking
potential.  We assume both $\psi$ and $\phi$ have mass of order
the soft SUSY breaking scale of order 1 TeV (where we will need
to test the consistency of this assumption). We assume that the
potential for the $\psi$ field gives a positive mass squared at
the origin, while the $\phi$ field has negative mass squared at
the origin. Furthermore, we assume that the cosmological constant
is zero at the minimum of both $\psi$ and $\phi$. The specific
form of the potential we choose is
\begin{equation}
V=M^4 \cos^2\left({\phi/\sqrt{2} f } \right)+{m_{\psi}^2\over 2} \psi^2
+{\psi^4\phi^2+\phi^4 \psi^2 \over 8 {M'}^2}\ ,
\end{equation}
where we have taken the scalar field to be real. When the
parameters are motivated by supersymmetry breaking, we refer to
our models by the name supernatural inflation. We will see that
one very naturally obtains the correct magnitude of density
perturbations, and sufficiently many e-foldings of inflation,
using parameters and a potential which are well motivated in
supersymmetric models.

The $\psi$ mass is $m_\psi$ and the magnitude of the
{(imaginary)} $\phi$ mass term (at the origin) is $m_{\phi}\equiv
M^2/f$.  During inflation, $\phi$ is confined near the origin.
The field $\psi$ slowly rolls towards the origin and inflation
ends about when $\psi=\psi_c= \sqrt{2M'm_{\phi}}$.  It will turn
out that either $m_\psi/m_\phi$ or $M'/M_p$ is small, so that
during inflation the term $m_\psi^2 \psi^2$ is small relative to
$M^4$.  The Hubble parameter during inflation is therefore
approximately $H=\sqrt{8 \pi/3}M^2/M_p$. 

We expect $f$ is of order $M_P$, or equivalently, $m_\phi^2$ is
of order $m_{3/2}^2$.  Although it looks like we took a very
special form for the $\phi$-potential, the use of the cosine is
not essential.  As can be seen from a Taylor expansion, only at
the very late stages of inflation are terms other than the
constant and mass term relevant. We could equally well have
specified a potential which is truncated at fifth order in the
fields, or which has different higher order terms. Although both
$\psi$ and $\phi$ might be moduli or standard model flat
direction fields, we assume their potentials are of very
different form; the particular case we assume is illustrative of
how a model could work. 

The constraint from density perturbations in the slow-roll regime
is \cite{spectrum,cobe} ${V^{3/2}/[{\tilde M_p^3 (dV/ d\psi)}]}=6
\times 10^{-4}$, where $\tilde M_p \equiv M_p/\sqrt{8\pi}$. This
gives the constraint
\begin{equation} \label{form}
{M^5 \over m_\psi^2 M_p^3}\sqrt{f \over M'} e^{\mu_\psi^2
N/3}=6.7 \times 10^{-6}\ .
\end{equation}
Here we have defined $\mu_\psi=m_\psi/H$ and have measured time
in $e$-foldings away from the time $N=0$ when $\psi=\psi_c$
(where inflation ends at positive $N$).  It is clear that a lower
$M'$ makes the value of $\psi$ at the end of inflation lower,
which in turn {\it increases} the density perturbations. The
exponential in Eq.~(\ref{form}) determines the scale dependence
of the density perturbations, characterized by the scalar index.

The scalar index $\alpha_s$ is readily determined from the scale
dependence of the density perturbations to be $-\mu_\psi^2/3$. 
This can be seen directly from the formula for density
perturbations above. Alternatively, it is extracted from the
general formula \cite{turner}
\begin{equation}
n=1-2\alpha_s=1-3\left({V'\over V}\right)^2+2 {V'' \over V}\ ,
\end{equation}
where dimensionful factors should be compensated by $\tilde M_p$.
Notice that the second term is negligible for all models for
which the inflaton field value is much less than $M_p$.  This is
readily seen from the fourth expression in Eq.~(\ref{neqn}),
which implies $\tilde M_p V'/V\approx \Delta \phi/ (N M_p)$. The
third term is positive in our model, because the inflaton field
rolls toward, rather than away, from the origin during the end of
inflation.

We see for this model that $n$ is always greater than 1, and is
very close to 1 for small $\mu_\psi$, which is the case for large
$M'$.  This differs from the usual prediction for new inflation
or chaotic inflation models.  The current upper bound on $n$ is
uncertain as is summarized in Ref. \cite{cgl}.  These bounds,
along with the validity of slow roll, prevent too large values of
$\mu_\psi$.  It turns out that in this model, the constraint
for the correct magnitude of density fluctuations requires that
when $M'\approx M_p$, $\mu_\psi\sim 1/100$, when $M'\approx M_G$,
$\mu_\psi \sim 1/10$, and when $M'\approx M_I$, $\mu_\psi \sim
1$. These constraints are given in more detail in Ref. \cite{us}.

Another distinctive feature of these models is that the ratio of
the tensor to scalar contribution to the quadrupole ${\cal R}={T
/ S}\sim \left({V'/ V}\right)^2 \approx 0$. Again this follows
from the small value of the inflaton field $\psi$ near the end of
inflation.

As we have argued in the first section, models of inflation which
have only a single field should have the inflaton field taking a
value of order $M_p$ near the end of inflation if 50 e-foldings
are to be obtained without fine tuning.  The combination of
negligible ${\cal R}$ and $n$ never below 1 are distinctive
features of these models which should help distinguish them from
other possible inflationary models in the future.

Another distinctive feature of the perturbation spectrum from
this model will be a spike at small wavelengths. This spike can
be calculated from a detailed study of the evolution of the
$\psi$ and $\phi$ fields, as is performed in Ref. \cite{us}. Here
we give the qualitative features of the evolution.

As we have emphasized, the $\psi$ field acts as a trigger to end
inflation. While the $\phi$ field is confined to the origin, the
evolution of the $\psi$ field is straightforward; it moves in
towards the origin according to the slow roll equation of motion.

While the $\psi$ field is big, the $\phi$ field has a large mass,
so the field is strongly confined to the origin.  As $\psi$ rolls
in towards the origin, the $\phi$ mass is decreasing.  Once the
$\phi$ mass is sufficiently small, it random walks away from the
origin due to de Sitter fluctuations.  As the mass squared of the
$\phi$ field becomes negative, the field value becomes
sufficiently large that the classical potential dominates over
the de Sitter fluctuations as a driving force.  The $\phi$ field
then moves according to its classical equations of motion towards
its true minimum. 

Because of the coupling of the $\psi$ and $\phi$ fields, the mass
of the $\psi$ field increases as $\phi$ increases. Eventually,
the time dependent $\psi$ mass is sufficiently large that the
$\psi$ field begins to oscillate about its true minimum as a
coherent state, and subsequently decay. Once the $\phi$ field
oscillates about its true minimum, the exponential expansion of
the universe ceases, and ultimately the $\phi$ field decays,
permitting the universe to reheat to a temperature somewhat above
the weak scale.

The spike in the spectrum arises from fluctuations in the $\phi$
field during the stage in its evolution where de Sitter
fluctuations drive its motion.  We have calculated this spike
\cite{us} and constrain our model so that inflation ends
sufficiently rapidly that the spike is only relevant to
perturbations on small (as yet unobservable) scales, less than 1
Mpc. This has given us a lower bound on $\mu_\phi$. We have also
checked that the magnitude of the spike is consistent with known
black hole constraints on these small scales.

An alternative model can be constructed based on a renormalizable
potential. Although this seems strange since we are considering
flat direction fields, it is often the case, even in the MSSM,
that fields cannot be simultaneously flat.  So we take the
potential to contain the soft supersymmetry breaking terms as
before but to contain a Yukawa coupling involving $\psi$ and
$\phi$. Specifically
\begin{equation}
V=M^4 \cos^2\left({\phi/\sqrt{2} f } \right)+{m_{\psi}^2\psi^2
\over 2} + \lambda^2 {\psi^2 \phi^2 \over 4}\ .
\end{equation}

This model has the essential features of the FDHI model of the
previous section. The difference is the value of $\psi_c$ which
in this model is $\psi_c={\sqrt{2} m_\phi / \lambda}$. The
density fluctuations give the constraint
\begin{equation}
{\lambda H^3 e^{\mu_\psi^2 N/3} \over m_\psi^2 m_\phi}=1.6 \times
10^{-4} \ .
\label{eq:18}
\end{equation}
If $\lambda$ is of order unity, to satisfy Eq.~(\ref{eq:18})
requires $\mu_\phi>10^3$. However, if $\lambda\approx
10^{-4}-10^{-5}$, the model works perfectly with $\mu_\psi$ and
$\mu_\phi$ both of order unity.  We see there is virtually no
fine tuning, so long as a small $\lambda$ exists.  In
Ref.~\cite{us} we discuss examples, and construct an explicit GUT
extension of the MSSM in which the small Yukawa coupling required
by the particle physics provides the small coupling needed for
the correct amplitude of density fluctuations. 

\section{Conclusions}

We have shown that with more than one field it is possible to
construct models of inflation with no small parameters. 
Furthermore, the mass scales which seem to most naturally appear
in these models are of order $m_{3/2}$, about 1 TeV, and $M_I$,
about $10^{11}$ GeV, leading to a natural association with
supersymmetric models.  These models give rise to the correctly
normalized density perturbations, even though the Hubble constant
is quite low, of order $10^3-10^4$ GeV, because the value of the
inflaton field at the end of inflation is much lower than the
Planck scale.  The key to producing more such models is a
sensitive dependence of the $\phi$ potential on the value of the
$\psi$ field, so that the motion of the $\psi$ field can trigger
the end of inflation while its value is small.

An important feature which distinguishes these models from
previous examples of hybrid inflation is that both the $\psi$ and
$\phi$ fields are assumed to have small mass.  This is the more
natural choice if both fields correspond to flat direction
fields. Furthermore, because supersymmetry is broken during
inflation, it is difficult to maintain the hierarchy between the
two fields due to large radiative corrections.  One would need
further complications to maintain this hierarchy naturally.  The
small mass of $\phi$ leads to the characteristic spike on small
length scales.

Our model is also consistent with the gravitino bound and Affleck
Dine baryogenesis \cite{us}. 

It seems that multifield models are probably the most natural
models which can implement inflation with weak scale Hubble
constant, and that furthermore, these are probably the most
natural inflation models in that they involve no new small
parameters. The requisite small parameters arise naturally from
the ratio of mass scales. These models have the further
advantages that they can be explicitly realized and one can
calculate the relevant parameters for any particular
implementation. They might even occur in simple extensions of the
MSSM. 

Perhaps the most important property of a model is its
testability, and our proposed models have several characteristics
that are in principle observable.  The scalar index $n$ which
characterizes the scale dependence of density perturbations is
always greater than unity.  It is very close to unity for the
model of Eq.~(8) with $M'$ at the Planck or GUT scale, but for
$M'$ at the intermediate scale or for the model of Eq.~(11), it
could be as large as 1.2 for the parameters
presented in our plots.  In all cases tensor perturbations are
negligible.  An especially distinctive feature is a large spike
in the density perturbation spectrum at present wavelengths of
about 1 Mpc or less.

\acknowledgments
We are very grateful to Andrew Liddle, David Lyth, and Ewan
Stewart for discussions and the Aspen Center for Physics where
these discussions took place.  We also thank Sean Carroll, Csaba
Cs\'aki, Arthur Kosowsky, Andrei Linde, and Bharat Ratra for
their comments.  We thank Bernard Carr, Jim Lidsey, Avi Loeb,
Paul Schechter, and Paul Steinhardt for discussions and
correspondence about black holes. We also thank Krishna Rajagopal
for his comments on the manuscript.

The work of L.R. was supported in part by the Department of
Energy under cooperative agreement DE-FC02-94ER40818, NSF
grant PHY89-04035, NSF Young Investigator Award, Alfred Sloan
Foundation Fellowship, and DOE Outstanding Junior Investigator
Award.  The work of M.S. was supported in part by MIT
\hbox{UROP}.  The work of A.H.G. was supported in part by the
Department of Energy under cooperative agreement
DE-FC02-94ER40818.

\end{document}